\documentclass[prc,showpacs,twocolumn,aps]{revtex4}

\usepackage{bm}
\usepackage{amsmath}
\usepackage{epsfig}
\usepackage{graphics}
\usepackage{graphicx}

\begin{document}


\title{Single particle spectra based of modern effective interactions} 

\author{C.~Barbieri}
  \thanks{Present address: GSI, Planckstr. 1, 64291, Darmstadt, Germany}
  \email{C.Barbieri@gsi.de}
  \affiliation
     {Gesellschaft f\"ur Schwerionenforshung, Planckstr. 1,
     64291, Darmstadt, Germany \\  }
  \affiliation
     {TRIUMF, 4004 Wesbrook Mall, Vancouver,
          British Columbia, Canada V6T 2A3 \\  }


\date{\today}

\begin{abstract}
 The self-consistent Green's function method is applied to ${}^{16}$O
using a G-matrix and $V_{UCOM}$ as effective interactions, both 
derived from the Argonne $v_{18}$ potential.
 The present calculations are performed in a larger model space than
previously possible. The experimental
single particle spectra obtained with the G-matrix
are essentially independent of the oscillator length of the basis.
 The results shows that $V_{UCOM}$ better reproduces spin-orbit splittings
but tends to overestimate the gap at the Fermi energy.
\end{abstract}
\pacs{ 24.10.cn, 21.60.-n, 21.30.Fe, 21.60.Jz}

\maketitle

 A fundamental problem in nuclear physics is how to obtain descriptions of
finite nuclei starting from a microscopic nuclear Hamiltonian.
Much progress has been achieved for few body systems. The Green's function
Monte Carlo~\cite{GFMC} technique is able to give exact results up to A=12,
while the no-core shell model~\cite{NCSH} is accurate for even larger nuclei.
 A wide range of exact methods is also available for very light
systems~\cite{fb_benchm}. 
 Other recent attempts to push the limits of {\em ab initio} methods
into the medium mass region have focussed on the nucleus of ${}^{16}$O and
its neighbor isotopes~\cite{Fujii,CCMprl04}. These gave predictions
for separation energies and spin orbit splitting of the orbits near the
Fermi level.
Coupled cluster theory appears to produce converged results
for these nuclei~\cite{CCMjul05}.
 These achievements have been possible by computing the contributions
of long-range correlations (LRC) directly within very large models spaces
where, however, one still needs to employ a proper effective
interaction that accounts for the excluded degrees of freedom.
 In particular the effects due to short-range correlations (SRC) can be
separated efficiently by such partitioning procedure, since they are
characterized by high momenta degrees freedom~\cite{DiBa04}.

 In {\em ab initio} methods the effective interaction is to be
computed microscopically.
 Typically, two classes of approaches are possible to derive
effective interactions from a standard realistic nucleon-nucleon force.
 Block-Horowitz theory makes use of the Feshbach projection formalism
to devise an energy dependent interaction~\cite{BlHo,Luu}.
 This gives solutions for every eigenstate with non zero
projection onto the model space, however, the energy dependence severely
complicates the calculations.
 The G-matrix interaction~\cite{Martino_PR} is a first order approximation
to the Block-Horowitz scheme.
 Alternatively, one can employ a proper unitary transformation to map a finite
set of solutions of the initial Hamiltonian into states belonging to the
model space. In this case, one has the advantage to work with an energy
independent interaction.
  Examples of such approach are the Lee-Suzuki method~\cite{LeeS},
V$_{low-k}$~\cite{Vlowk} and
the unitary correlator operator method (UCOM)~\cite{Vucom98,Vucom03,Vucom04}.
The UCOM formalism is such that one can apply the inverse transformation
to reinsert SRC into the nuclear wave function.
%
 A discussion of the similarities and differences between Lee-Suzuki
and Block-Horowitz is given in Ref.~\cite{ByronEPJA}.
 These methods, in principle, generate effective many-nucleon forces
in addition to the two-nucleon (2N) interaction. 
In practice, however, calculating medium and large nuclei one is forced
to work with only 2N Hamiltonians (or at most with weak 3N forces).
 It is therefore important to investigate how this truncation affects the
results using the different approaches outlined above.

In Ref.~\cite{BaDi01} we proposed to employ a set of Faddeev equations
within the self-consistent Green's function (SCGF) approach~\cite{DiBa04}
to obtain a microscopical description of LRC. This allows to couple
simultaneously quasiparticles to both particle-hole (ph) and
particle-particle/hole-hole (pp/hh) collective excitations.
 Such formalism was later applied to ${}^{16}$O to investigate
mechanisms that could possibly quench the spectroscopic factors of mean field
orbits~\cite{BaDi02}. These calculations were already performed in
a no-core fashion. However, the model space employed was still somewhat limited
and phenomenological corrections were applied to tune the values of specific
single particle (sp) energies 
(that allows to study correlations by artificially suppressing
the couplings among selected excitation modes).
Note that here and in the following we use the terms {\em sp energies} and
{\em sp spectra} to refer to the poles of the one-body Green's function
(defined below Eq.~(\ref{eq:g1})~). These represent the excitation energies
of the A$\pm$1 neighbor nuclei, which are observable quantities.
In this letter the calculations of Ref.~\cite{BaDi02} are repeated by avoiding
any phenomenology and employing a large model space.
 We discuss the results of 2N interactions belonging
the two types discussed above, namely a standard G-matrix and $V_{UCOM}$.

\begin{figure}
 \begin{center}
      \includegraphics[width=2.4in]{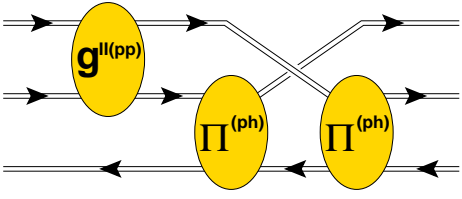}
\caption[]{Example of a Feynman diagram included in the all-order summation
generated by the set of Faddeev equations.
Double lines represent the dressed one-particle Green's function $g(\omega)$,
which propagates quasiparticles (rightward arrows) and quasiholes (leftward
arrows). The ellipses propagate collective excitations of the
nucleus~(Eqs.~(\ref{eq:Pi}) and~(\ref{eq:g2})~).
\label{FaddSum} } 
\end{center}
\end{figure}

We consider the calculation of the sp Green's function
\begin{equation}
 g_{\alpha \beta}(\omega) ~=~ 
 \sum_n  \frac{ \left( {\mathcal X}^{n}_{\alpha} \right)^* \;{\mathcal X}^{n}_{\beta} }
                       {\omega - \varepsilon^{+}_n + i \eta }  ~+~
 \sum_k \frac{ {\mathcal Y}^{k}_{\alpha} \; \left( {\mathcal Y}^{k}_{\beta} \right)^* }
                       {\omega - \varepsilon^{-}_k - i \eta } \; ,
\label{eq:g1}
\end{equation}
from which both the one-hole and one-particle spectral functions, for the
removal and addition of a nucleon, can be extracted.
In Eq.~(\ref{eq:g1}),
${\mathcal X}^{n}_{\alpha} = {\mbox{$\langle {\Psi^{A+1}_n} \vert $}}
 c^{\dag}_\alpha {\mbox{$\vert {\Psi^A_0} \rangle$}}$
~(${\mathcal Y}^{k}_{\alpha} = {\mbox{$\langle {\Psi^{A-1}_k} \vert $}}
 c_\alpha {\mbox{$\vert {\Psi^A_0} \rangle$}}$) are the
spectroscopic amplitudes for the excited states of a system with
$A+1$~($A-1$) particles and the poles $\varepsilon^{+}_n = E^{A+1}_n - E^A_0$
~($\varepsilon^{-}_k = E^A_0 - E^{A-1}_k$) correspond to the excitation
energies with respect to the $A$-body ground state.
The one-body Green's function can be computed by solving the
Dyson equation~\cite{diva2005,fetwa}.
\begin{equation}
 g_{\alpha \beta}(\omega) =  g^{0}_{\alpha \beta}(\omega) \; +  \;
   \sum_{\gamma \delta}  g^{0}_{\alpha \gamma}(\omega) 
     \Sigma^*_{\gamma \delta}(\omega)   g_{\delta \beta}(\omega) \; \; ,
\label{eq:Dys}
\end{equation}
where the irreducible self-energy $\Sigma^*_{\gamma \delta}(\omega)$ acts
as an effective, energy-dependent, potential that governs
the single particle behavior of the system.
The self-energy is expanded in a Faddeev series as in Fig.~\ref{FaddSum},
which couples the exact propagator $g_{\alpha \beta}(\omega)$ (which is itself
a solution of Eq.~(\ref{eq:Dys})~) to other phonons in the system~\cite{BaDi01}. 
The relevant information regarding ph and pp/hh collective excitations
is included in the polarization and the two-particle propagators.
Respectively,
\begin{eqnarray}
 \Pi_{\alpha \beta , \gamma \delta}(\omega) &=& 
 \sum_{n \ne 0}  \frac{  {\mbox{$\langle {\Psi^A_0} \vert $}}
            c^{\dag}_\beta c_\alpha {\mbox{$\vert {\Psi^A_n} \rangle$}} \;
             {\mbox{$\langle {\Psi^A_n} \vert $}}
            c^{\dag}_\gamma c_\delta {\mbox{$\vert {\Psi^A_0} \rangle$}} }
            {\omega - \left( E^A_n - E^A_0 \right) + i \eta } 
\nonumber \\
 &-& \sum_{n \ne 0} \frac{  {\mbox{$\langle {\Psi^A_0} \vert $}}
              c^{\dag}_\gamma c_\delta {\mbox{$\vert {\Psi^A_n} \rangle$}} \;
                 {\mbox{$\langle {\Psi^A_n} \vert $}}
             c^{\dag}_\beta c_\alpha {\mbox{$\vert {\Psi^A_0} \rangle$}} }
            {\omega - \left( E^A_0 - E^A_n \right) - i \eta } \; ,
\label{eq:Pi}
\end{eqnarray}
and
\begin{eqnarray}
 g^{II}_{\alpha \beta , \gamma \delta}(\omega) &=& 
 \sum_n  \frac{  {\mbox{$\langle {\Psi^A_0} \vert $}}
                c_\beta c_\alpha {\mbox{$\vert {\Psi^{A+2}_n} \rangle$}} \;
                 {\mbox{$\langle {\Psi^{A+2}_n} \vert $}}
         c^{\dag}_\gamma c^{\dag}_\delta {\mbox{$\vert {\Psi^A_0} \rangle$}} }
            {\omega - \left( E^{A+2}_n - E^A_0 \right) + i \eta }
\nonumber \\  
&-& \sum_k  \frac{  {\mbox{$\langle {\Psi^A_0} \vert $}}
    c^{\dag}_\gamma c^{\dag}_\delta {\mbox{$\vert {\Psi^{A-2}_k} \rangle$}} \;
                 {\mbox{$\langle {\Psi^{A-2}_k} \vert $}}
                  c_\beta c_\alpha {\mbox{$\vert {\Psi^A_0} \rangle$}} }
            {\omega - \left( E^A_0 - E^{A-2}_k \right) - i \eta } \; ,
\label{eq:g2}
\end{eqnarray}
which describe the one-body response and the propagation
of two-particles/two-holes.
In this work, $\Pi(\omega)$~and $g^{II}(\omega)$ are obtained by solving
the dressed RPA~(DRPA) equations~\cite{schuckbook,DRPApaper},
which account for the redistribution of strength in the sp spectral function.
Since this information is carried by the correlated
propagator~$g_{\alpha \beta}(\omega)$, Eq.~(\ref{eq:Dys}),
the SCGF formalism requires an iterative solution.
 It can be proven that full self-consistency guarantees to satisfy the
conservation of the number of particles and other basic quantities~\cite{BaKa}.

The coupled cluster studies of Ref.~\cite{CCM05prl,CCMjul05} found that
eight major harmonic oscillator shells can be sufficient
to obtain converging results for ${}^{16}$O with G-matrix interactions.
At the same time, the experience with the calculations of Ref.~\cite{BaDi02}
suggests that high partial waves do not contribute sensibly.
In this work, all the orbits of the first eight shells with orbital angular
momentum $l \le$~4 were included.
 Inside this model space a G-matrix and the $V_{UCOM}$
potential were employed as effective interactions.
The former was computed using the CENS library routines~\cite{Martino_PR,cens}.
For the latter, the UCOM matrix-elements code~\cite{VucomME} was employed
with the constraint $I_\vartheta=0.09$~fm$^3$. This choice of the UCOM
correlator reproduces, in perturbation theory, the binding energies of several
nuclei up to ${}^{208}$Pb~\cite{Vucom_Oct05}.
In both cases the Argonne $v_{18}$ potential~\cite{WS_av18} was used as
starting interaction. However, we chose to neglect the Coulomb and
the other charge independence breaking terms (i.e., to set $\alpha =~$0)
in the present work.
In applying the SCGF formalism, the Hartree-Fock (HF) equations
(Brueckner-Hartree-Fock (BHF) for the G-matrix) where first
solved for the unperturbed propagator $g^{(B)HF}_{\alpha \beta}(\omega)$.
This was employed in the first iteration.
 After that, the (dressed) solution $g_{\alpha \beta}(\omega)$ was used
to solve the DRPA for $\Pi(\omega)$~and $g^{II}(\omega)$ and then
the Faddeev equations for an improved self-energy. 
 In each iteration the two most important fragments close to the Fermi level
were retained for each partial wave, both in the quasiparticle and
the quasihole domains.
 The remaining strength do not affect sensibly the sp states that will be
discussed below and was collected, at each iteration, according to the
corresponding (B)HF orbitals.
In the present work, calculations were iterated until reaching convergence
(to within 200~keV) for the sp energies nearby the Fermi level.
With the available computational resources, self-consistence was achievable
for these states except in the case of the G-matrix with harmonic oscillator
length $b_{HO}\le$ 1.8 fm (as discussed below).
To test the iteration procedure we computed the total number
of particles obtained with $V_{UCOM}$. The first calculation, based on
the HF propagator, gives A=16.4 (a 2.5\% error).
At self-consistency 15.99$<$A$<$16.02 (due to numerical errors), showing
the adequacy of our approach.
More details on the SCGF/Faddeev formalism are given in
Ref.~\cite{BaDi01,BaDi02,DiBa04}.
As already noted, however, no phenomenological corrections were applied 
in the present work.

\begin{figure}[ht]
 \begin{center}
\includegraphics[height=4.5cm]{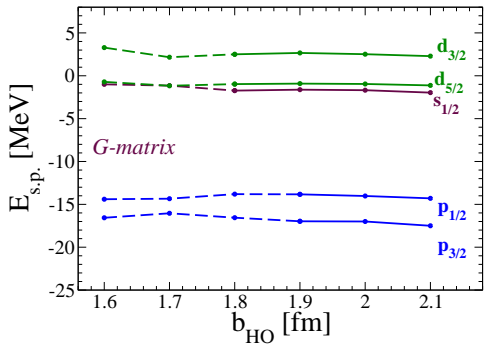}
 \end{center}
\caption[]{
  Single particle spectrum obtained with the G-matrix as a function
 of the oscillator length. 
  The dashed lines refer to values of $b_{HO}$ for which self-consistence
  was not reached.}
\label{Gvsb}
\end{figure}

\begin{figure}
 \begin{center}
\includegraphics[height=4.5cm]{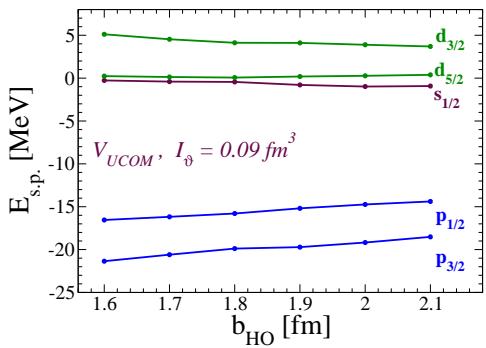}
 \end{center}
\caption[]{
  Self-consistent single particle spectrum obtained with $V_{UCOM}$
 as a function of the oscillator length. }
\label{Vvsb}
\end{figure}

 Calculations have been performed for oscillator lengths in the interval
$b_{HO}$=1.6--2.1~fm.
The sp spectra obtained at self-consistency are shown in Figs.~\ref{Gvsb}
and~\ref{Vvsb}.
In the case of a G-matrix, these orbitals appear independent, within numerical
accuracy, of the oscillator frequency for b$_{HO}\ge$~1.9~fm.
 Below this the trend is similar. However, the structure of the quasiparticle
spectral function becomes increasingly complex for the $l$=1 waves
that correspond to the 
$pf$ shell.
These break in more than two relevant fragments and their sp energies depend
strongly on $b_{HO}$, suggesting a non negligible contamination of the c.o.m.
motion.
In this situation the above approach for obtaining self-consistency
cannot be carried out reliably. We show the result of a typical iteration
by dashed lines in Fig.~\ref{Gvsb} but discard these results in the following
discussion.
This complication does not happen for $V_{UCOM}$, which is a softer
interaction and generates, for each shell, at most one main fragment
and a smaller satellite peak near the Fermi energy.
 Fig.~\ref{Vvsb} shows that the spin orbit splittings for this interaction 
are approximately constant, although the sp energies are not yet
independent of the oscillator length.
 This can be understood considering that these spin orbit partners correspond
to particularly simple and similar configurations (one particle or one hole
on top of the correlated ground state). Conversely, separation energies
are linked to the to total binging energy of neighbor isotopes.
Larger model spaces will probably be required for a full convergence with $V_{UCOM}$.

\begin{table}
\begin{center}
\begin{tabular}{lcccccccccccc}
\hline
\hline
 b$_{HO}$[fm]~=              &~&  1.6 &~&  1.7 &~&  1.8 &~&  1.9 &~&  2.0 &~&  2.1 \\
\hline
$G-matrix$: \\
$\Delta E_{p_{1/2}-p_{3/2}}$ & &   -  & &   -  & &   -  & &  3.1 & &  3.1 & &  3.2 \\
$\Delta E_{d_{3/2}-d_{5/2}}$ & &   -  & &   -  & &  3.5 & &  3.6 & &  3.5 & &  3.4 \\
~\\
V$_{UCOM}$: \\
$\Delta E_{p_{1/2}-p_{3/2}}$ & &  4.7 & &  4.4 & &  4.1 & &  4.5 & &  4.4 & &  4.1 \\
$\Delta E_{d_{3/2}-d_{5/2}}$ & &  4.9 & &  4.4 & &  4.1 & &  3.9 & &  3.6 & &  3.3 \\
 \hline \hline
\end{tabular}
\end{center}
\caption[]{Spin orbit splittings (in MeV).}
\label{sosplit}
\end{table}
The splittings obtained from both interactions are reported in 
Tab.~\ref{sosplit}. For the $0p$ orbits these are practically constant for all
the oscillator lengths. 
In general the G-matrix predicts about a half of the experimental value.
Better solutions are obtained with the present choice of the UCOM correlator.
 For the $0d$ orbits the results for the two interactions are more similar
to each other but not totally independent of the oscillator length.

The same type of LRC studied here (which are predominantly of 2p1h and 2h1p
in character) were also considered in Ref.~\cite{Fujii}. There, the effective
interaction was derived in the unitary-model-operator approach and an explicit
diagonalization was performed. The resulting sp energies and spin-orbit
splitting, however, showed a stronger dependence on the oscillator length than
the one found in this work.
The spectrum obtained in Fig.~\ref{Gvsb} is nearly convergent, suggesting that
the all order summation employed here and the proper accounting of the
fragmentation of the sp strength allow to select relevant
configurations beyond the bare 2p1h/2h1p level.
Coupled cluster calculations are also available for ${}^{16}$O with
an analogous $v_{18}$/G-matrix~\cite{CCMjul05}. These authors report
splittings of the $0p$ and $0d$ orbits larger than those of Tab.~\ref{sosplit}
by about~1.5 and 0.5~MeV, respectively.
 The present work includes LRC only in the form of coupling to small amplitude
excitations of the core ---which can be described at the DRPA level.
More complex collective modes are also present~\cite{BaDi03} and should be
included for a full solution of the many-body problem.
 We note, for example, that the phenomenological studies of Ref.~\cite{BaDi02}
suggest further contributions to the $p_{3/2}$ quasihole wave function coming
from couplings to the first excited 0$^+$ state in oxygen.
 Testing this conjecture would first require being able to reproduce
the correct excitation energy of this level ---since it can couple effectively
only when it is low enough in energy.
To our knowledge this is still a challenge for the available {\em ab initio}
methods.

\begin{figure}
 \begin{center}
 \includegraphics[height=4.cm]{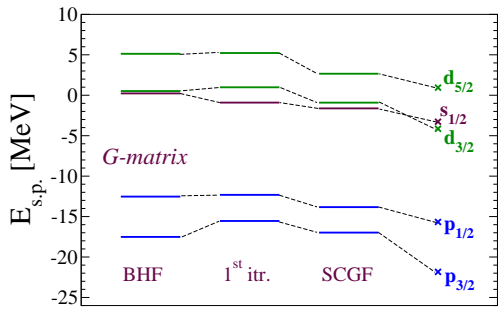}
 \end{center}
 \caption[]{
 Single particle spectra generated by the G-matrix in the HF approximation,
after the first iteration and in the full self-consistent solution.
 The points on the left represent the experimental values for the O17/O15 case.}
 \label{Gmtx_spe} 
\end{figure}

\begin{figure}
 \begin{center}
\includegraphics[height=4.cm]{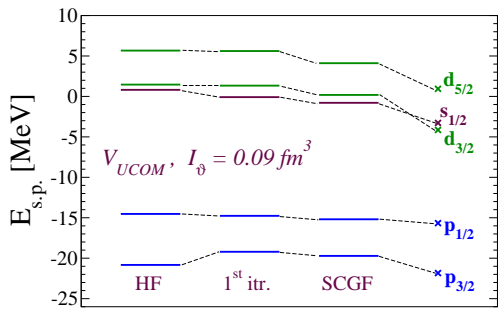}
 \end{center}
\caption[]{
 Same as Fig.~\ref{Gmtx_spe} but for the $V_{UCOM}$ interaction. }
 \label{Vucom_spe} 
\end{figure}

Figures~\ref{Gmtx_spe} and~\ref{Vucom_spe} show the effects of LRC
on the sp spectrum for $b_{HO}$=~1.9~fm, and compare to the experimental
values for the addition/removal of a neutron.
 For both interactions the coupling to collective phonons reduces the
splitting of the $0p$ orbits, with respect to the HF approximation.
 Including the effects of fragmentation tends instead to compress the $sd$
shell and to lower the whole spectrum.
 The self-consistent results for the energy gap between particle and hole
states,
$\Delta E_F = \varepsilon_{d_{5/2}} - \varepsilon_{p_{1/2}}$, are 13.0~MeV
with the G-matrix and 15.4~MeV with $V_{UCOM}$. Both of them exceed the
experimental value of 11.5~MeV.
 The mean square radii obtained are r$_{rms}$=2.63~fm~(G-matrix)
 and r$_{rms}$=2.45~fm~($V_{UCOM}$).

  Both the UCOM and the renormalization group approaches have the capability
of shifting the binding energies for A=3,4 systems along
the Tjon line~\cite{Nogga:2004ab,Vucom_May05}.
 In the first case this can be achieved by varying the correlator in the
tensor-isoscalar channel. For $V_{low-k}$, one modifies the momentum cut off.
Usually, tuning the binding energies to the experimental value increases
the non locality of the interaction and leads improved spin orbit splittings,
as seen in Tab.~\ref{sosplit}.
 On the other hand, our $V_{UCOM}$ result for $\Delta E_F$ --with 2N forces--
overestimates the experiment, more than the G-matrix does.
 This behavior is seen already at the HF level for soft interactions
like $V_{UCOM}$ and $V_{low-k}$~\cite{Paar:2005py,Vucom_Oct05} and it is only
slightly modified by the LRC considered here.
 In all cases, three-body forces appear necessary in order to reproduce
the whole spectrum of observations. We note, however, that the UCOM method
offers some advantages to reduce the contribution needed from many-body forces
since it allows to treat SRC in different channels separately~\cite{Vucom04}.

In conclusion, SCGF calculations have been performed for the first time
in a large model space, including up to eight oscillator shells.
Long-range correlations in the form of coupling sp to ph and pp/hh
RPA modes were investigated for ${}^{16}$O. A comparison was made
between the results of a G-matrix and the $V_{UCOM}$ effective interactions,
both derived from same realistic potential (Argonne $v_{18}$).
The spectra of adjacent nuclei were found to be nearly convergent
for the G-matrix, while they depend only weakly on the oscillator
length for $V_{UCOM}$.
In general it was found that the LRC effects considered here, tend to compress
the spectra of A$\pm$1 nuclei but do not affect sensibly the
the gap between quasiparticle and quasihole energies at the Fermi level.

\acknowledgments
The author is in debt with B.~K.~Jennings for suggesting this work,
as well as with H.~Feldmeier and R.~Roth for several useful discussions.
This work was supported in part by the Natural Sciences and Engineering
Research Council of Canada (NSERC). Part of the calculations were performed
at the TRIUMF/UBC node of WestGrid Canada. 



\end{document}